
\font\mbf=cmbx10 scaled\magstep1

\def\bs{\bigskip}
\def\ms{\medskip}
\def\np{\vfill\eject}

\def\ni{\noindent}
\def\cl{\centerline}

\def\title#1{\cl{\mbf #1}\ms}
\def\ctitle#1{\bs\cl{\bf #1}\par\nobreak\ms}
\def\stitle#1{\bs{\ni\bf #1}\par\nobreak\ms}

\def\ref#1#2#3#4{#1\ {\it#2\ }{\bf#3\ }#4\par}
\def\refb#1#2#3{#1\ {\it#2\ }#3\par}
\def\ANY{Ann.\ N.Y.\ Acad.\ Sci.}
\def\CQG{Class.\ Qu.\ Grav.}
\def\PR{Phys.\ Rev.}
\def\RNC{Riv.\ Nuovo Cimento}

\def\abs#1{\left\vert#1\right\vert}
\def\O#1{\left.#1\right\vert_S}
\def\T#1{\left.#1\right\vert_T}

\def\I{\int_S\mu}
\def\D{{\cal D}}
\def\L{{\cal L}}
\def\R{{\cal R}}
\def\l{\Lambda}

\def\k{\kappa}
\def\t{\theta}
\def\s{\sigma}
\def\o{\omega}

\def\i{\lim_{\xi\to\infty}}
\def\half{{\textstyle{1\over2}}}
\def\third{{\textstyle{1\over3}}}
\def\quart{{\textstyle{1\over4}}}
\def\sixth{{\textstyle{1\over6}}}

\magnification=\magstep1

\title{A cosmological constant limits the size of black holes}
\ctitle{Sean A. Hayward,* Tetsuya Shiromizu \& Ken-ichi Nakao}
\cl{Department of Physics}
\cl{Kyoto University}
\cl{Kyoto 606-01}
\cl{Japan}
\ms\cl{*Faculty of Mathematical Studies}
\cl{University of Southampton}
\cl{Southampton SO9 5NH}
\cl{United Kingdom}
\bs\cl{Revised 7th March 1994}
\bs\ni{\bf Abstract.}
In a space-time with cosmological constant $\l>0$
and matter satisfying the dominant energy condition,
the area of a black or white hole cannot exceed $4\pi/\l$.
This applies to event horizons where defined,
i.e.\ in an asymptotically deSitter space-time,
and to outer trapping horizons (cf.\ apparent horizons) in any space-time.
The bound is attained if and only if the horizon
is identical to that of the degenerate `Schwarzschild-deSitter' solution.
This yields a topological restriction on the event horizon,
namely that components whose total area exceeds $4\pi/\l$ cannot merge.
We discuss the conjectured isoperimetric inequality
and implications for the cosmic censorship conjecture.
\np
\stitle{I. Introduction}\ni
Recent studies of black holes
in space-times with positive cosmological constant $\l$
have revealed some surprising differences compared with
black holes in the zero $\l$ case.
Kastor and Traschen have found a family of solutions to
the Einstein-Maxwell equations
which describe an arbitrary number of charged black holes
in an otherwise closed cosmos [1].
For the zero $\l$ case,
these solutions reduce to the static Majumdar-Papapetrou solutions,
but for positive $\l$,
the black holes collide and coalesce for a certain range of parameters,
providing the first exact solutions describing coalescing black holes.
In a previous study
of uncharged multi-black-hole initial data,
another remarkable result was discovered, and proposed as a general feature:
black holes whose total mass is larger than a certain critical value
do not coalesce, i.e.\ no new apparent horizon forms [2].
Also, the area of the apparent horizon that forms in the subcritical case
has an upper bound, approached as the critical case is approached.
This led to the conjecture that $\l$ yields an upper bound
on the area of a black hole,
with the bound $12\pi/\l$ being established for maximal slices [3].
A similar study has since been made of the Kastor-Traschen solutions,
with the same conclusions:
there is a critical mass beyond which the black holes do not coalesce,
and when they do coalesce,
there is an upper bound on the area of the new apparent horizon [4].
In this article,
we show that there is a sharp bound $4\pi/\l$
on the area of a black or white hole.
\ms
We emphasize that by `black or white hole' we mean
any region of trapped surfaces
that has a boundary of the outer type,
according to the classification in [5].
The boundary, referred to as the {\it trapping horizon} [5],
replaces the foliation-dependent and less general concept of
apparent horizon [6],
and henceforth we will consistently speak of trapping horizons
rather than apparent horizons.
It transpires that the area bound also applies to event horizons,
but the event horizon, like the apparent horizon,
is defined only in asymptotically flat space-times,
or for positive $\l$, in asymptotically deSitter space-times [3],
a rather special case.
\ms
As a preliminary,
we examine exact solutions and other examples in Section II.
The area bound is established for outer trapping horizons in Section III,
and for event horizons in Section IV.
In the Conclusion, we discuss the conjectured isoperimetric inequality
and implications for the cosmic censorship conjecture.

\stitle{II. Examples}\ni
Carter [7] showed that the `Kerr-Newman' black-hole solutions
could be generalized to include a cosmological constant $\l$.
In particular, there is the analogue of the Reissner-Nordstr\"om solution,
parametrized by mass $M$, charge $Q$ and $\l$, with line-element
$$ds^2=-F\,dT^2+F^{-1}dR^2+R^2dS^2\qquad
F(R)={Q^2\over{R^2}}-{2M\over R}+1-\third\l R^2\eqno(1)$$
where $dS^2$ refers to the unit 2-sphere.
The global structure of these solutions is described
for the $Q=0$ case in [8] and [2], and in general in [9].
For a certain range of $(M,Q,\l)$ and choice of topology,
the solutions describe a charged black hole of the Wheeler wormhole type,
existing in an asymptotically deSitter cosmos.
The trapping horizons are given by the zeros of $F(R)$,
and have area $4\pi R^2$.
Consider two special cases, as follows.
\item{(i)} The `Schwarzschild-deSitter' case, $Q=0$.
Fixing $\l$, the trapping horizons occur at $M=\half R-\sixth\l R^3$,
as depicted in Figure 1(i).
For $M<\sqrt{1/9\l}$ there are two horizons,
namely a black-hole horizon and a cosmological horizon,
which coincide in the degenerate $M=\sqrt{1/9\l}$ case.
Note that the area of the black-hole horizon does not exceed $4\pi/\l$,
and attains $4\pi/\l$ in the degenerate case.
\item{(ii)} $\abs{Q}=M$.
The trapping horizons occur at $M=R\pm\sqrt{\l/3}R^2$,
as depicted in Figure 1(ii).
For $M<\sqrt{3/16\l}$ there are three horizons,
namely inner and outer black-hole horizons and a cosmological horizon,
with the last two coinciding in the degenerate $M=\sqrt{3/16\l}$ case.
The area of the outer black-hole horizon does not exceed $3\pi/\l$,
attained in the degenerate case.
Note that adding charge has reduced the maximal area.
\par\ni
Checking the full range of $(M,Q,\l)$ confirms that
the area of the outer black-hole horizon does not exceed $4\pi/\l$
for any of these solutions.
That the same bound should apply to any black hole
is suggested on thermodynamic grounds.
The entropy of a black hole is essentially its area,
and one would expect entropy to be maximized
in the most symmetric configuration,
namely the stationary, spherically symmetric `Schwarzschild-deSitter' solution.
A similar argument motivates the conjectured isoperimetric inequality
in asymptotically flat space-times [10,11].
\ms
Another test of the area bound is provided by multi-black-hole solutions
where the black holes collide and coalesce.
We use `coalescence' to mean the appearance of a new region of trapped surfaces
outside the original trapping horizons of the incoming black holes.
Thus a double trapping horizon forms,
with inner and outer horizons both enclosing the original trapping horizons,
as depicted in Figure 60 of [6].
If such a coalescence occurs,
one may ask whether the new outer trapping horizon satisfies the area bound.
There are two known examples for $\l>0$.
\item{(iii)}
The initial data for the vacuum Einstein equations found by Nakao et al [2],
which develop into a multi-black-hole cosmos.
For the case of two equal masses,
it was found that there is a critical total mass $\sqrt{1/9\l}$
such that coalescence occurs in the subcritical but not supercritical case.
In Figure 13 of [2], it can be seen that the area of the new horizon
is bounded above by a value which appears to be $4\pi/\l$.
So the collision seems to miraculously preserve the area bound.
\item{(iv)} The Kastor-Traschen multi-black-hole solutions,
which are exact Einstein-Maxwell solutions
describing charged $\abs{Q}=M$ black holes [1].
Again, for the case of two equal masses,
it has been found that there is a critical total mass,
approximately $\sqrt{3/16\l}$,
such that
coalescence occurs in the subcritical but not supercritical case [4].
For these solutions,
the area of the new trapping horizon is bounded above by $3\pi/\l$.
As for the single-mass case (ii),
the lower maximum may be attributed to the charge on the holes.
\par\ni
This is a very curious phenomenon.
It seems that nature is conspiring to prevent
the formation of too large a black hole.
Attempting to create a large black hole by colliding smaller ones
either produces a black hole satisfying the bound
or does not work at all---the black holes refuse to coalesce.
\np
\stitle{III. Trapping horizons}\ni
We adopt a recent approach to trapping horizons,
replacing apparent horizons,
which has elucidated some of their fundamental properties [5].
One of the main outcomes of this analysis
is that it is crucial to distinguish between
inner and outer trapping horizons,
since they have quite different properties---for instance,
outer trapping horizons are generically spatial
while inner trapping horizons are generically Lorentzian.
We shall see that another such difference is the area bound,
which applies to outer but not inner trapping horizons.
\ms
We recall the basic definitions and notation of [5].
Consider a double-null foliation,
i.e.\ two foliations of null 3-surfaces
labelled by coordinates $\xi_+$ and $\xi_-$,
intersecting in a foliation of spatial 2-surfaces $S$.
Introduce the normal 1-forms $n_\pm=-\hbox{d}\xi_\pm$
and the dual vectors $N_\pm=g^{-1}(n_\pm)$,
where $g$ is the space-time metric.
Then $g(N_\pm,N_\pm)=0$ since the 3-surfaces of constant $\xi_\pm$ are null.
Introduce the normalization $\hbox{e}^f=-g(N_+,N_-)$
and the induced 2-metric $h=g+2\hbox{e}^{-f}n_+\otimes n_-$.
The area form, Ricci scalar and covariant derivative of $h$ are denoted by
$\mu$, $\R$ and $\D$ respectively.
The area of a compact 2-surface is $$A=\I.\eqno(2)$$
The evolution vectors $u_\pm=\partial/\partial\xi_\pm$
are assumed future-pointing,
and are related to the null normals by
$u_\pm-r_\pm=\hbox{e}^{-f}N_\mp$,
where the shift 2-vectors $r_\pm=h^{-1}h(u_\pm)$.
Denoting the Lie derivatives along $u_\pm-r_\pm$ by $\L_\pm$,
the expansions $\t_\pm$, shears $\s_\pm$,
inaffinities $\nu_\pm$
and anholonomicity (or twist) $\o$ are defined by
$$\eqalignno{
&\t_\pm=\half h^{cd}\L_\pm h_{cd}&(3a)\cr
&\s^\pm_{ab}=h_a^ch_b^d\L_\pm h_{cd}
-\half h_{ab}h^{cd}\L_\pm h_{cd}&(3b)\cr
&\nu_\pm=\L_\pm f&(3c)\cr
&\o_a=\half\hbox{e}^{-f}h_{ab}[N_+,N_-]^b.&(3d)\cr}
$$
Defining $\phi_+=T(u_+-r_+,u_+-r_+)$ and $\rho=T(u_+-r_+,u_--r_-)$
in terms of the material energy tensor $T$,
the relevant components of the Einstein equation are
the {\it focussing} and {\it cross-focussing} equations,
which are respectively
$$\eqalignno{
&\L_+\t_++\nu_+\t_++\half\t_+^2+\quart\s^+_{ab}\s_+^{ab}=-8\pi\phi_+&(4a)\cr
&\L_-\t_++\t_+\t_-+\hbox{e}^{-f}(\half\R-\tau_a\tau^a-\D_a\tau^a)
=8\pi\rho+\hbox{e}^{-f}\l&(4b)\cr}
$$
where it is convenient to introduce $\tau=\o-\half\D f$.
The dominant energy condition implies
$$\eqalignno{
&\phi_+\ge0&(5a)\cr
&\rho\ge0.&(5b)\cr}
$$
Note that all the results of [5] also apply to the case $\l>0$,
since the $\l$-term in the Einstein equation, regarded as a material source,
satisfies the dominant energy condition.
\ms
A {\it marginal surface} is a spatial 2-surface $S$
on which one null expansion vanishes,
fixed henceforth as $\O{\t_+}=0$.
A {\it trapping horizon} is the closure $\overline{T}$ of a 3-surface $T$
foliated by marginal surfaces on which
$\T{\t_-}\not=0$ and $\T{\L_-\t_+}\not=0$,
where the double-null foliation is adapted to the marginal surfaces.
The trapping horizon and marginal surfaces are said to be
{\it outer} if $\T{\L_-\t_+}<0$,
{\it inner} if $\T{\L_-\t_+}>0$,
{\it future} if $\T{\t_-}<0$
and {\it past} if $\T{\t_-}>0$.
For a future outer horizon, the idea is that
the outgoing light rays are diverging just outside the horizon,
and converging just inside,
and that the ingoing light rays are converging.
The {\it trapping gravity} $\k$ of an outer trapping horizon is defined by
$$4\k^2=\T{-\hbox{e}^f\L_-\t_+}.\eqno(6)$$
An example is provided by the Carter black-hole cosmos of Section II,
for which the event horizon is an outer trapping horizon,
with the Cauchy horizon and cosmological horizon being inner trapping horizons.
In general, the outer trapping horizon may be taken as the definition
of the outer boundary of a black or white hole.
Note that one cannot use the event horizon
as a general definition of the boundary of a black hole,
since it is defined only if conformal infinity exists,
i.e.\ in asymptotically flat or deSitter space-times.
\ms
Marginal surfaces which lie in trapping horizons must be orientable,
due to the differing signs of $\O{\t_+}$ and $\O{\t_-}$.
In the context of gravitational collapse,
one normally expects the marginal surfaces to be compact.
The non-compact case can also be treated,
provided certain asymptotic conditions are satisfied,
namely integrability of $\R$, $\D_a\tau^a$ and $\sqrt{\tau_a\tau^a}$
over the surface.
Such a marginal surface is said to be {\it well adjusted}.
\ms\ni
{\it Theorem 1.}
In a space-time with cosmological constant $\l>0$
and matter satisfying the dominant energy condition,
a well adjusted, future or past, outer marginal surface
has finite area $A$ bounded above by $A<4\pi/\l$.
\ms\ni
{\it Proof.}
On an outer marginal surface $S$,
the cross-focussing equation (4b) gives
$$-4\k^2+\O{(\half\R-\tau_a\tau^a-\D_a\tau^a)}=\O{8\pi\hbox{e}^f\rho}+\l.
\eqno(7)$$
Since $\k^2>0$, the dominant energy condition implies
$0<\l<\O{\half\R}-\O{\D_a\tau^a}$.
Now $\I\D_a\tau^a=0$ by the Gauss theorem
and $\I\R\le4\pi\chi$ by the Cohn-Vossen inequality,
where $\chi$ is the Euler-Poincar\'e characteristic of $S$,
with $\chi=2$ for a sphere, $\chi=1$ for a plane,
and $\chi\le0$ for any other orientable 2-manifold.
Integrating the inequality over $S$, $$0<A\l<2\pi\chi\eqno(8)$$
so that the area $A$ is finite, whence $S$ is compact.
Since $\chi>0$, $S$ has spherical topology.
Also $A\l<4\pi$.
\ms\ni
In particular, this means that black and white holes
lie within the Hubble radius $\sqrt{3/\l}$,
which corresponds to an area $12\pi/\l$.
Actually, there is a more precise result in terms of
the irreducible energy $m$,
angular energy $a$, material energy $q$
and root-mean-square trapping gravity $k$, defined by
$$\eqalignno{
&16\pi m^2=A&(9a)\cr
&4\pi a^2=m^2\I\tau_a\tau^a&(9b)\cr
&8\pi q^2=A\I\hbox{e}^f\rho&(9c)\cr
&Ak^2=\I\k^2.&(9d)\cr}
$$
\ms\ni
{\it Theorem 2.}
In a space-time with cosmological constant $\l>0$
and matter satisfying the dominant energy condition,
the irreducible energy $m$,
angular energy $a$, material energy $q$
and r.m.s.\ trapping gravity $k$
of a well adjusted, future or past, outer marginal surface satisfy
$${A\l\over{4\pi}}=1-{a^2\over{m^2}}-{q^2\over{m^2}}-16m^2k^2.\eqno(10)$$
\ms\ni
{\it Proof.}
Integrate the cross-focussing equation (4b)
using the Gauss-Bonnet theorem for $S$ with spherical topology, $\I\R=8\pi$.
\ms\ni
The formula shows explicitly how the area of a black hole is reduced by
angular momentum, matter and trapping gravity.
If any one of these quantities is non-zero,
there is a lower effective bound on the area.
\ms
So far, we have restricted the marginal surfaces to be non-degenerate,
$\k^2>0$, in which case the bound is not attained.
This may be relaxed to $\k^2\ge0$.
The additional cases, namely where $\k^2$ vanishes somewhere,
are referred to as {\it degenerate} marginal surfaces.
\ms\ni
{\it Theorem 3.}
In a space-time with cosmological constant $\l>0$
and matter satisfying the dominant energy condition,
a well adjusted, degenerate outer trapping horizon
has area $4\pi/\l$ if and only if
it is identical to that of the degenerate `Schwarzschild-deSitter' solution.
\ms\ni
{\it Proof.}
On a well adjusted, degenerate outer marginal surface $S$,
the cross-focussing (4b) equation integrates to
$${A\l\over{4\pi}}\le
\half\chi-{a^2\over{m^2}}-{q^2\over{m^2}}-16m^2k^2\eqno(11)$$
using the Cohn-Vossen inequality again.
Now $A\l=4\pi$ requires $\chi\ge2$,
so that $S$ has spherical topology again, $\chi=2$.
Also $a$, $q$ and $k$ must vanish.
Hence $\O\tau$, $\O\rho$ and $\O\k$ also vanish,
which in turn means that $\O\R=2\l$, so that $S$ is metrically spherical.
A trapping horizon $T$ foliated by such constant-area spheres must be null,
by the area theorem for trapping horizons [5].
Constant area also implies that the internal expansion $\T{\t_+}$ vanishes,
and the focussing equation (4a) shows that
$\T{\s_+}$ and $\T{\phi_+}$ also vanish.
This has now exhausted the free data on a null 3-surface [12],
and the data is identical to that of the horizon of
the degenerate `Schwarzschild-deSitter' solution,
i.e.\ $T$ can be embedded in this solution.
Note that a uniqueness result for the whole space-time
would require additional initial data to be specified off the horizon.
\ms\ni
Finally, we note that it is crucial for Theorems 1--3
that the trapping horizon or marginal surface be outer rather than inner.
For inner marginal surfaces,
the trapping-gravity term in the cross-focussing equation
has the opposite sign,
and if large enough can swamp the angular and material terms
to yield marginal surfaces of any topology and arbitrarily large area.
As an example,
the cosmological horizon of the `Schwarzschild-deSitter' solution
has area between $4\pi/\l$ and $12\pi/\l$.
It has been conjectured that in an asymptotically deSitter space-time,
the cosmological horizon,
defined as the boundary of the past domain of dependence of $\Im^+$,
should have area less than or equal to $12\pi/\l$, the deSitter value [13].
Although this conjecture is still open,
it does not generalize to inner trapping horizons in arbitrary space-times,
since an inner marginal surface can have arbitrarily large area.
As an example, the innermost horizon of the `Schwarzschild-deSitter' solution
has arbitrarily large area
in the case where there are naked singularities instead of black holes.

\stitle{IV. Event horizons}\ni
The classical work on black holes
is mainly concerned with the idealization of asymptotic flatness [6].
For $\l>0$, there is a similar class of asymptotically deSitter space-times.
Conformal infinity $\Im^\pm$ is spatial rather than null,
but many aspects of the space-times can be treated by methods
which are well known in the asymptotically flat case.
In particular, one may define the event horizon
as the boundary of the causal past of $\Im^+$.
Since the area of the event horizon is non-decreasing [3],
and the event horizon is expected
to approach the trapping horizon(s) asymptotically,
it seems that the area bound should also apply to the event horizon.
A preliminary step is to establish that
connected sections of the event horizon
approach outer marginal surfaces as $\Im^+$ is approached.
\ms\ni
{\it Lemma.}
Consider the event horizon $H$
of a strongly future asymptotically predictable,
asymptotically flat or asymptotically deSitter space-time.
Foliate $H$ by spatial 2-surfaces,
with future-pointing null normals $u_\pm-r_\pm$,
with $u_+-r_+$ tangent to the event horizon,
and $\xi$ an affine parameter along $u_+-r_+$.
Then
$$\eqalignno{
&\i\t_+(H)=0&(12a)\cr
&\i\L_-\t_+(H)\le0&(12b)\cr}
$$
assuming the weak energy condition or null convergence condition [6].
\ms\ni
{\it Proof.}
First recall that $\t_+(H)\ge0$ [3,6].
The focussing equation (4a) yields the inequality
$\t_+\le(1/\t_0+(\xi-\xi_0)/2)^{-1}$,
from which it follows that $\i\t_+(H)=0$.
Consider a connected component $S$ of $\overline{H}\cap\overline{\Im^+}$,
and a neighbouring foliation of $\Im^+$
containing another 2-surface $S'\subset\Im^+$.
Propagating $u_\pm-r_\pm$ from $S$ to $S'$,
construct the null 3-surface $H'$ through $S'$ in the $u_+-r_+$ direction,
in some neighbourhood $U$ of $\Im^+$.
Suppose there is a $p\in H'\cap U$ such that $\t_+(p)<0$.
Then there is a neighbourhood $V$ of $p$ in $H'\cap U$ such that $\t_+(V)<0$,
and a compact set $C\subset S''\cap V$,
where $S''$ is a spatial 2-surface of the foliation.
This leads to a contradiction, since it is impossible
that $C\subset\hbox{int}[J^-(\Im^+)]$,
by the same argument as in
the proof of the area theorem for event horizons [3,6].
Thus $\t_+(H'\cap U)\ge0$,
and so $\i\L_-\t_+(H)\le0$.
\ms\ni
{\it Theorem 4.}
In a strongly future asymptotically predictable,
asymptotically deSitter space-time
with matter satisfying the dominant energy condition,
the area $A$ of a connected section of the event horizon is bounded above by
$A\le4\pi/\l$,
with equality if and only if the event horizon
is identical to that of the degenerate `Schwarzschild-deSitter' solution.
\ms\ni
{\it Proof.}
Integrating the cross-focussing equation (4b)
over connected sections $S_\xi$ of the event horizon and taking the limit,
it follows from (12) and the Cohn-Vossen inequality that
$$\i A_\xi\l\le\i2\pi\chi_\xi\eqno(13)$$
whence $\i A_\xi$ is finite and so $\i S_\xi$ is compact.
Since the area of the event horizon is non-decreasing [3],
$A_\xi\le\i A_\xi$, it follows that $\i A_\xi>0$, and so $\i\chi_\xi>0$.
Since the event horizon is orientable, $\i\chi_\xi=2$.
In particular,
$$A_\xi\le\i A_\xi\le4\pi/\l.\eqno(14)$$
Attaining the bound on the whole event horizon $H$ requires constant area,
so $\t_+(H)=0$.
This means that the event horizon is also a trapping horizon,
and Theorem 3 then shows that the bound is attained
if and only if $H$ is identical to
the horizon of the degenerate `Schwarzschild-deSitter' solution.
\ms\ni
{\it Corollary.}
Asymptotically, the event horizon consists of sections of spherical topology.
\ms\ni
Note that the bound applies to each connected section of the event horizon.
The total area of all the connected sections at a given time
may exceed the bound.
In such a case,
there is a topological restriction on the development of the event horizon,
namely that contemporaneous sections whose total area exceeds $4\pi/\l$
cannot subsequently combine into a single connected section.
In other words, such black holes cannot merge in this sense.
This is exactly what is observed in
the examples of Kastor and Traschen [1,4] and Nakao et al [2].

\stitle{V. Conclusion and discussion}\ni
In asymptotically flat space-times, it is well known that
the event horizon cannot divide and has non-decreasing area [6].
This is also true in an asymptotically deSitter space-time~[3],
but we have shown that there are additional restrictions:
the area of each connected section of the event horizon
cannot exceed $4\pi/\l$,
and so disconnected components cannot merge
if their total area exceeds this value.
\ms
This has an interesting consequence for the cosmic censorship conjecture.
Consider initial data representing two or more black holes,
defined in the sense of outer trapping horizons,
such that the total area exceeds $4\pi/\l$.
Suppose further that these black holes subsequently collide.
Since the resulting space-time cannot contain an event horizon,
one might conjecture
that the outcome of such a collision would be a naked singularity [14].
This would be a blatant violation of the cosmic censorship conjecture
in either weak [15] or strong [16] form.
What actually happens for the examples of Kastor and Traschen and Nakao et al
is rather remarkable [2,4].
In the so-called subcritical case, the black holes coalesce,
producing a new asymptotically stable trapping horizon
whose area can be arbitrarily close to the bound.
If the black holes collided to form an asymptotically stable object
in the supercritical case as well,
that object would have to be a naked singularity,
assuming some sort of `no-hair' theorem
to the effect that a stable solution must be one of the Carter solutions.
Instead,
it turns out that the black holes refuse to collide in the supercritical case,
and simply keep their distance.
Physically, one could ascribe this phenomenon
to the repulsive effect of the cosmological constant.
This lends considerable moral support to the cosmic censorship conjecture
in the following sense:
the delicate balance between gravitational attraction and cosmic repulsion
tips exactly when the black holes can no longer coalesce,
and a collision would be in danger of producing a naked singularity.
\ms
The inequality $A\le4\pi/\l$ bears a striking resemblance
to the Penrose-Gibbons isoperimetric inequality
$$A\le16\pi M_0^2\eqno(15)$$
which is conjectured to relate
the ADM mass $M_0$ and the area $A$
of the outermost trapping horizon (or apparent horizon) $T$
of an asymptotically flat space-time,
assuming matter satisfying the dominant energy condition [10,11].
Equality is conjectured to be attained only for the Schwarzschild solution.
The conjecture may be strengthened to the following chain of inequalities:
$$
A\le\lim_{T\to i^+}A
\le\lim_{\Im^+\to i^+}16\pi M^2
\le16\pi M^2
\le\lim_{\Im^+\to i^0}16\pi M^2
\le16\pi M_0^2\eqno(16)
$$
where $M$ denotes the Bondi-Sachs mass.
The first inequality follows from the area theorem for trapping horizons [5],
and the third and fourth from the Bondi-Sachs mass-loss result.
The two missing links are plausible,
since $M$ and $M_0$ are both limits of the same quasi-local energy $E$ [17],
and $16\pi E^2/A\to1$ as $T\to i^+$.
If this chain of inequalities is correct,
and there exists an asymptotically flat initial data set
on which $A>16\pi M_0^2$,
this would also be a blatant violation of the cosmic censorship conjecture.
Again, one may try to construct such a counterexample
by trying to create a large black hole from the collisions of smaller ones.
In numerical examples, the isoperimetric inequality is preserved,
providing further moral support for the cosmic censorship conjecture [11].
\ms
Although formulated for the $\l=0$ case,
the isoperimetric inequality is analogous to the area bound
in that it gives a limit on the size of black holes,
regarding the ADM mass as fixed.
It also gives a necessary condition for a surface to be marginal.
This raises the question of whether a necessary and sufficient condition
for the existence of a trapped surface can be given in terms of initial data.
For a light-cone, this is indeed possible [18].
For a spatial 3-surface, there is no such general result,
nor even a precise conjecture.
At the heuristic level, there is the hoop conjecture [19].
The light-cone result is independent of $\l$,
but it is not clear how $\l$ should affect the hoop conjecture.
\ms
Both asymptotically flat and asymptotically deSitter space-times
are rather special,
and since the event horizon is defined only in such cases,
the general study of black holes
must instead be concerned with the trapping horizon.
The fundamental properties of trapping horizons
have only recently been established [5].
In particular, the area of a future outer trapping horizon is non-decreasing
and, as we have shown, cannot exceed $4\pi/\l$.
Given that a black-hole collision
involves the formation of a new outer trapping horizon,
one would also like to know if the area of the new horizon
is necessarily greater than the sum of the areas of the original horizons.
If this is true, this yields tests of the cosmic censorship conjecture
which can be formulated in arbitrary space-times.
Research in this direction is in progress.
\bs\ni
Acknowledgements.
TS and KN would like to thank H.~Sato, K.~Maeda, H.~Kodama and M.~Sasaki
for their helpful discussion and criticism.
SAH thanks the Southampton relativity group for hospitality.
\bs
\begingroup
\parindent=0pt\everypar={\global\hangindent=20pt\hangafter=1}\par
{\bf References}\ms
\ref{[1] Kastor D \& Traschen J 1993}\PR{D47}{5370}
\ref{[2] Nakao K, Yamamoto K \& Maeda K 1993}\PR{D47}{3203}
\ref{[3] Shiromizu T, Nakao K, Kodama H \& Maeda K 1993}\PR{D47}{R3099}
\refb{[4] Shiromizu T, Nakao K \& Hayward S A 1993}
{Horizons of the Kastor-Traschen multi-black-hole cosmos}{(in preparation)}
\ref{[5] Hayward S A 1993}
{General laws of black-hole dynamics, \PR}D{(to appear)}
\refb{[6] Hawking S W \& Ellis G F R 1973}
{The Large-Scale Structure of Space-Time}{(Cambridge University Press)}
\refb{[7] Carter B 1973 in}{Black Holes}
{ed: DeWitt C \& DeWitt B S (Gordon \& Breach)}
\ref{[8] Gibbons G W \& Hawking S W 1977}\PR{D15}{2738}
\ref{[9] Brill D R \& Hayward S A 1994}\CQG{11}{359}
\ref{[10] Penrose R 1973}\ANY{224}{125}
\refb{[11] Gibbons G W 1984 in}
{Global Riemannian Geometry}{ed: Willmore \& Hitchin N J (Ellis Horwood Ltd)}
\ref{[12] Hayward S A 1993}\CQG{10}{779}
\ref{[13] Boucher W, Gibbons G W \& Horowitz G T 1984}\PR{D30}{2447}
\ref{[14] Brill D R, Horowitz G T, Kastor D \& Traschen J 1994}\PR{D49}{840}
\ref{[15] Penrose R 1969}\RNC{1}{252}
\refb{[16] Penrose R 1979 in}{General Relativity, an Einstein Centenary Survey}
{ed: Hawking~S~W \& Israel W (Cambridge University Press)}
\ref{[17] Hayward S A 1994}\PR{D49}{831}
\ref{[18] Hayward S A 1992}\CQG9{L115}
\refb{[19] Thorne K S 1972 in}{Magic without Magic}{ed: Klauder (Freeman)}
\endgroup

\stitle{Figure captions}\ni
1. Location of the trapping horizons for the Carter $(M,Q,\l)$ solutions,
for (i) $Q=0$ and (ii) $\abs{Q}=M$.
The branches of the curves correspond to the cosmological horizon $C$
and the outer black-hole horizon (or event horizon) $O$,
meeting at the degenerate horizon $D$,
and in the case of (ii),
the inner black-hole horizon (or Cauchy horizon) $I$.

\bye